# Unstable dynamics of model vicinal surfaces:
# Initial and intermediate stages

F. Krzyżewski[1], M. Załuska-Kotur[1], A. Krasteva[2], H. Popova[3] and V. Tonchev [4, 1] [*]

[1]Institute of Physics, Polish Academy of Sciences, al. Lotników 32/46, 02-668 Warsaw, Poland.
[2]Institute of Electronics, Bulgarian Academy of Sciences, 72 Tzarigradsko chaussee blvd, 1784 Sofia, Bulgaria
[3]R. Kaischew Institute of Physical Chemistry, Bulgarian Academy of Sciences, Acad. G. Bonchev Str., block 11, 1113 Sofia, Bulgaria
[4] Faculty of Physics, Sofia University, 5 James Bourchier Blvd., 1164 Sofia, Bulgaria
[*]Corresponding author: tonchev@phys.uni-sofia.bg

**Abstract** We approach the oldstanding problem of vicinal crystal surfaces destabilized by step-down (SD) and step step-up (SU) currents from a unified modelling viewpoint with focus on both the initial and the intermediate stages of the instability. We reproduce and analyze the instability caused by the two opposite drift directions in the two fundamental situations of step motion – vicinal sublimation and vicinal growth. For this reason we develop further our atomistic scale model of vicinal crystal growth (Gr) destabilized by SD drift of the adatoms in order to account for also the vicinal crystal sublimation (Sbl) and the SU drift of the adatoms as an alternative mode of destabilization. In order to study the emergence of the instability we use the number of steps in the bunch (*bunch size*) $N$ as a measure and probe with small-size systems the model's stability against step bunching (SB) on a dense grid of points in the parameter space formed by the diffusion rate/step transparency, surface miscut and drift direction, for each of the four possible cases - Gr+SD, Gr+SU, Sbl+SD, Sbl+SU. The obtained stability diagrams show where the system is initially most unstable and provide a ground to study there the intermediate stages of the developed instability quantifying the surface self-similarity by the time-scaling of $N$. For each of the four enumerated cases we show that it reaches the universal curve $N = 2\sqrt{T/3}$, where $T$ is the time, properly rescaled with the model parameters. We confirm the value of the numerical pre-factor with results from a parallel study of models based on systems of ordinary differential equations (ODE) for the step velocity.

Step bunching (SB) is a phenomenon, archetypal for a class of surface instabilities - SB (of straight steps), step meandering (SM), simultaneous SB and SM, step bending, surface faceting. These and, with increasing share, their applied aspects and applications, were intensively studied in the years after 1990, in parallel with the dramatic development and sophistication of the experimental techniques for surface analysis, the available computational power and the ever-increasing demand from technological side. Although the first report on surface patterning under electric fields was reported as early as 1938 by Johnson [1] who studied tungsten filaments used in the incandescent lamps, the term *step bunching* itself appears in 50's ([2-4] and the references therein). Later the interest concentrated on almost flat on atomic scale tungsten surfaces [5] and the first theory of electromigration induced surface instability was advanced [6]. In a rather different context, the interest in SB is always closely connected with its, sometimes dramatic, impact on the processes of nanostructure layer-by-layer epitaxial growth as realized in various deposition techniques having their industrial realizations. Nowadays, the keyword *step bunching* bridges studies on material systems as diverse as $CH_3NH_3PbI_3$ [7], GaN [8], AlN [9], AlGaN [10], SiC [11-13], graphene [14, 15], PTCDA (perylene tetracarboxylic acid dianhydride)/Ag[16], pyronaridine/heme [17], $SrRuO_3$/(001) $SrTiO_3$[18], KDP [19-22], ferritin [23], etc. Models of step bunching are now part of novel approaches to exotic localizations such the sidewalls of the nanowires[24].

The onset of the modern stage of SB studies could be dated back to 1989 when Latyshev et. al [25] report the observation of the phenomenon on sublimating Si(111) surfaces destabilized by both step-up and step-down direct currents used to heat the surface but by only one of these for any fixed temperature. It was soon after hypothesized based on a simple Burton-Cabrera-Frank [26] (BCF) type of model(s) that it is the directional asymmetry (bias) in the diffusion of the charged surface atoms (adatoms) that makes the motion of the steps from the vicinal surface unstable[27-29]. As a result, the regular step distribution of the vicinal surface is changed and a surface that is sequence of groups of steps (bunches) separated by large terraces almost free of steps ("hills and valleys" structure) results. In the next years the phenomenon of SB was investigated actively by experimental techniques [30-34] analysed theoretically [35-47], and by numerical simulations [48-57]. It became clear only quite recently that SB can be caused at a given temperature by both current directions across the steps but on a metal surface – that of tungsten, W(110) [58]. Besides semiconductor and metal crystal surfaces also the insulator ones exhibit the same, adatom electromigration induced, type of behaviour [59].

The mechanisms of all enlisted above phenomena are far from being unified based on clear understanding of the interplay of surface elementary processes. In the narrower domain of SB induced by adatom electromigration the one caused by step-down flux of adatoms is easy to understand based on concepts within the BCF paradigm - the step velocity is obtained from the gradients at the steps of the diffusion field on a single terrace. Thus, when the contributions of the two terraces encircling the step are uneven the step motion could be unstable – when the contribution of the terrace that is behind the moving step is larger. This mechanism was used also to explain the step-up drift induced SB, for which change of the sign of adatom charge, the nature of the bias, etc., were evoked. The present paradigm was formed with the important input from S.Stoyanov [60] – the destabilization due to step-up adatom drift is following the emergence of a specific step property – the so called transparency (or permeability[61, 62]) – the adatoms, during their diffusion along the surface, almost do not feel the transparent steps as special points on the surface, thus not being sinks/sources of adatoms. This property allows for the adatom fluxes to carry information of the local ordering along the surface over many terraces across the steps [63]. None of these theories was able to comprise both mechanisms within a unified approach. Some other deficiencies of the existing paradigm are the unfinished program of finding the full set of scaling exponents/relations quantifying the intermediate stages of the instability, started by A. Pimpinelli et al. [37] and partially continued by J. Krug et al.[64] and the incorporation of the simultaneous step bunching and step meandering [65-67].

Developing further the recently introduced 1D atomistic scale model of vicinal surface growth destabilized by SD drift of adatoms[55, 56], a realization of fine-tuning of step transparency bound to the adatom diffusion and step kinetics, we are now able to reproduce the instability caused by the two opposite drift directions in the two fundamental situations of step motion – vicinal sublimation and vicinal growth.

Our model (vicCA) [55, 56] is a conceptual realization of vicinal surface, descending from left to right with adatoms on it and combines Monte Carlo (MC) with Cellular Automaton (CA) modules. It allows for quantitative analysis of the pattern formation dynamics by applying a modified monitoring protocol [39]. While the automaton rule acts in a parallel fashion providing simultaneous realization of all growth/sublimation events, in between the rule executions the adatom diffusion is simulated by the MC module which acts in a serial mode, dealing with adatom after adatom. The adatoms are stored in a separate table with surface concentration $c_0$ and diffuse on top of the vicinal surface without "feeling" the steps. Their diffusion is influenced by a jump asymmetry (bias), thus the jump probability $p_J$ in each direction (left, right) is dependent on $\delta$ – $p_J = (1/2 - \delta, 1/2 + \delta)$, with $|\delta| \leq 1/2$. Negative (positive) values of $\delta$ stand for bias pointing in the step-up (step-down) direction. Between two rule executions all the adatoms try on average $n_{DS}$ hops but only those that point at a neighbouring unoccupied lattice site are performed. When increasing $n_{DS}$ one departs from the diffusion-limited (DL) growth mode towards the kinetics-limited (KL) one and, simultaneously, increases the step transparency. Only the top atoms from the lattice are stored in an array through their corresponding heights, forming the vicinal stairway descending to the right with initial, vicinal distance between sequential steps of $l_0$. Crystal growth events are realized with the CA module which prescribes that an adatom right to a step sticks to it with a probability $p_G$, the step moves one lattice site to the right and the adatom is deleted from the table of adatoms. Only during the rule execution the adatoms enter into contact with the steps. In the present version of vicCA another part of the composite CA module rules the sublimation - a process opposite to the growth one and causing step motion to the left. Technically, the sublimation of the crystal surface is realized by checking whether there is an adatom on top of a site where single or macrostep is present. If not, adatom detachment is performed from the step with probability $p_S$, when detachment is accepted the lattice height at that site is reduced by one and new adatom is added to table with the adatoms at that position. In such a way we can simulate sublimation and growth as reversible processes. It is important to note that setting the attachment probability $p_G \leq 1$ at and detachment probability $p_S \leq 1$ from the steps is another way to increase step transparency even when $n_{DS}=1$.

Summarizing, a single time step of the vicCA model run starts with sublimation (growth) update and then $n_{DS}$ diffusional steps are made after which growth (sublimation) update is performed. The attachment (sticking) and detachment probabilities vary from 0 to 1. Finally, after all growth, diffusional and sublimation processes, adatom concentration is recovered to $c_0$. Thus, in the sublimation case the density of adatoms at the surface will exceed $c_0$ and the excess is deleted from the table with adatoms at random. In the growth case, when concentration of the adatoms is expired below $c_0$, we add necessary adatoms at random on the unoccupied sites thus maintaining the adatom concentration at $c_0$. More details of the above routine are available in [55, 56] and in the Supplement. In order to study the stability of this model against step bunching (SB) we scan the parameter space consisting of $c_0$, $n_{DS}$, $\delta$ and probabilities of adatom attachment (detachment) to (from) the steps, while fixing the initial vicinal distance $l_0$=15. Such distance has been chosen as quite small, so our simulations could be performed in finite time and at the same time not very small so system shows full spectrum of possible behaviour. Since in the model we do not incorporate a repulsion between steps and as a result the bunches formed consist of densely located single steps and also of macrosteps – steps of multiple height[68].

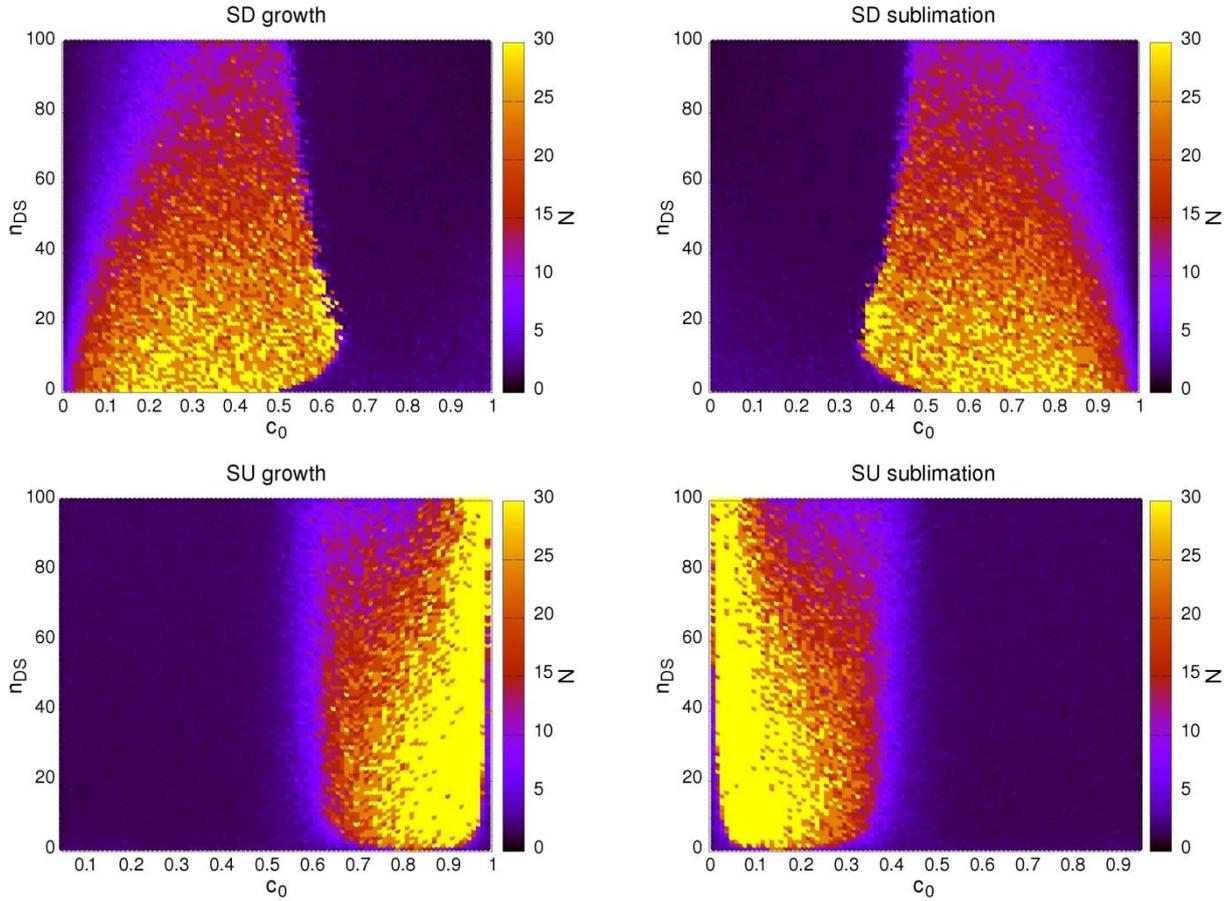

Figure 1. Stability diagrams of bunch size $N$ in the space of $n_{DS}$ and $c_0$ for the four investigated cases: growth destabilized by SD bias (a), sublimation - by SD bias (b), growth - by SU bias (c) and sublimation - by SU bias (d). Note the mirror symmetry between SD and SU case for given destabilization.

The stability diagram of the vicCA model is studied in the four possible combinations of vicinal process, growth and sublimation, and destabilizations – bias (drift) oriented step-down or step-up. In each point we start with a vicinal surface consisting of equally distributed steps. We start with fixed values of $\delta$=0.2 for the SD cases and $\delta$=-0.2 for the SU one, and perform runs each of them for 500 time units monitoring the average bunch size $N$. The results for the stability of the model are summarized in Figure 1. Panel (a), upper left, presents diagram of the mean bunch size $N$ as a function of $n_{DS}$ and $c_0$ for the case of irreversible growth ($p_G = 1$ and $p_S = 0$) with SD bias. In panel (*b*), upper right, we show the stability diagram of the sublimation destabilized by SD drift. This diagram possesses a mirror symmetry with respect to the *y*-axis with the previous plot – it corresponds to changing $c_0$ to (1-$c_0$). We have checked that SB is obtained only in such conditions for which rates of growth and sublimation are not too high. Those systems that grow or sublimate at high velocity do not form bunches but roughen. In order to ensure slow step motion and observe step bunching for the case of SU drift, one needs to make the process reversible – to realize both attachment and detachment of adatoms to/from the steps at each time step of the run. Thus for SU driven systems we assumed that $p_G$=1-$c_0$+0.1 and $p_S = c_0$ in the case of growing systems. When the system is sublimating we fix step detachment probability $p_S$=$c_0$+0.1 and set $p_G$=1-$c_0$. The results are shown in Figure 1c and Figure 1d. The areas where SB develops during growth and sublimation are separated. When crystal grows, Figure 1c, the most intensive SB is present at higher adatom concentrations varying between 0.75 and 1 and for broad range of $n_{DS}$ from ~10 to 100. For the sublimating system, Figure 1d, bunches emerge at lower concentrations ranging from 0 to 0.25 and the diagram is again as in the case of SD drift, a mirror image of Figure 1c.

It should be stressed that when switching from SD to SU bias the peak of the bunching instability shifts to the opposite corner of the diagram: from DL (small $n_{DS}$) to KL (large $n_{DS}$) and from $c_0$ to 1-$c_0$. Even if not perfect, such a symmetry is to be explained by mechanisms that are responsible for step bunching processes. One can also notice that the shape of parameter areas on the plot where SB is most intensive do not match the ones obtained for the opposite direction of the current. Hence in the present model, it is possible to obtain SB at SU and SD bias only after changing the other parameters of the simulation. The possible overlapping of SD and SU biased bunching can be examined by investigating the stability diagrams as a function of the drift direction. In Figure 2 such diagrams as a function of the concentration $c_0$ and value of the bias $\delta$ are presented. Positive (negative) values of $\delta$ cause step down (step up) drift of the adatoms. For opposite directions of bias and switching from growth to sublimation the model parametrisation (values of $p_G$ and $p_S$) is changed as described above. It is seen that step bunching occurs in the same system, growing or sublimating, for both bias directions, however the ranges of favourable surface densities (reflecting external flux of adatoms) are almost separated. There is only small range of $c_0$ values close to 0.4 where one finds SB with both drift directions.

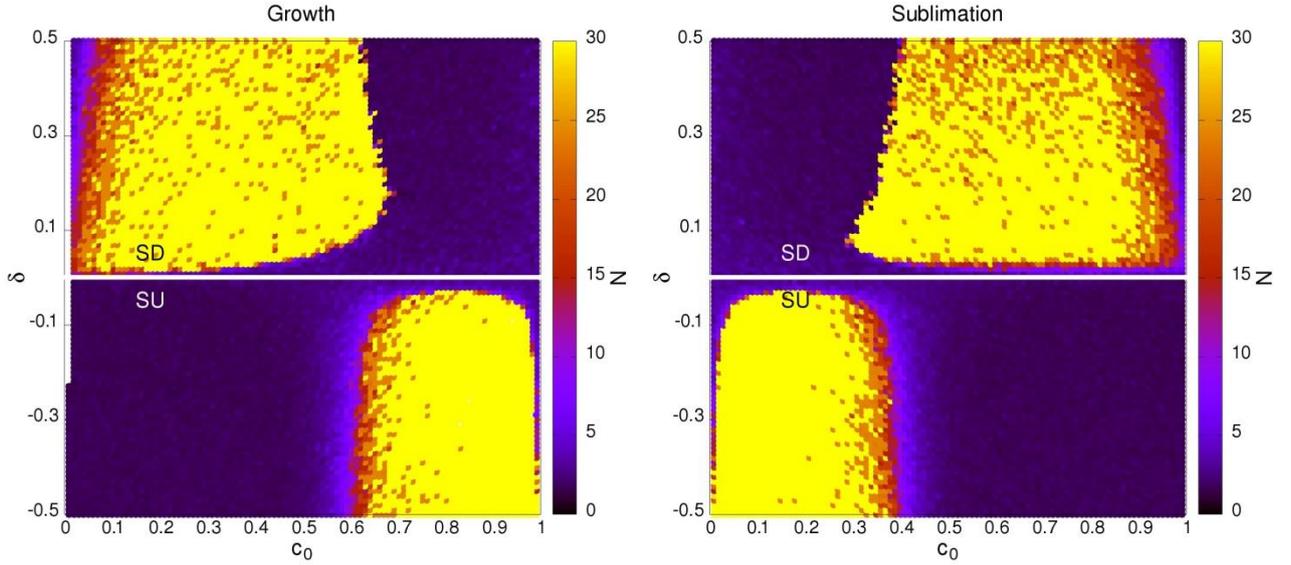

Figure 2 Stability diagram as a function of $c_0$ and $\delta$, $l_0$=15, $n_{DS}$=10. The positive $\delta$ values cause SD drift of adatoms.

As we can see in Figure 1 and Figure 2 the step bunching process is quite general phenomenon and happens for all combinations of growth/sublimation and step-up/step-down drifts. However the parameters for which the instability is observed are different. In general step-down drift induces bunching in DL/step impermeability regime, i.e. for low $n_{DS}$ values, whereas step-up drift should be combined with high $n_{DS}$ values, corresponding to KL/step permeability, in order to produce step bunching. Moreover, depending on combination between type of process and bias, low density or high density conditions are preferred and we discuss in more detail the instability development case by case. The first and the simplest one is growth (sublimation) at a low (high) density destabilized by bias in the SD direction and the step motion is controlled by the adatoms that stick (detach) to (from) the steps. Thus, only few diffusional hops are to be executed before each sticking (detachment) of an adatom to (from) a step and it is highly probable that each adatom (adatom vacancy) arriving at the step will be incorporated (cause a void) into the crystal. We can assume further that all adatoms (vacancies) that are present on a terrace eventually arrive at one of the step edges. With bias $\delta$ directed SD, $\delta$>0, the result is that the upper (lower) terrace contributes with more adatoms (vacancies) to the velocity $v_i$ of the $i$-th step than the lower (upper) one and one can write for the case of $n_{DS}$ = 1 when no contributions from the further located terraces are present:

$$v_i \equiv \frac{dx_i}{dt} = c\left(b_1 \Delta x_i + b_2 \Delta x_{i+1}\right) \qquad (1)$$

where is $\Delta x_i \equiv x_i - x_{i-1}$ is the length of the upper terrace, comprised by the steps at $x_i$ and $x_{i-1}$, $b_1$

and $b_2$ are (still unknown) functions of $\delta$ and $c = c_0$ ($c = -1 + c_0$) for growth (sublimation). A simplest linear stability analysis in which $\Delta x_i \to \Delta x_i + \partial x$ (and thus $\Delta x_{i+1} \to \Delta x_{i+1} - \partial x$) shows that any positive (negative) perturbation $\partial x$ will grow (decrease) further the step velocity and thus the length of $\Delta x_i$ provided $b_1 > b_2$ (for growth), thus resulting into the classical instability scenario – "these that have, will have more" (and vice versa), called in the general scientific context "Mathew effect" [69] after Matthew's 25:29. Since in the model step-step repulsions are not incorporated, there is no mechanism to prevent the formation of macrosteps. Beyond the initial destabilization caused by the uneven adatom fluxes from the two neigboring terraces that attach to (detach from) a step there is another mechanism of destabilization that comes into the play - when the lengths of the free from steps terraces between the dense step formations are increasing this leads to further destabilization because the hop asymmetry is amplified by the cumulation of many such hops before meeting a step.

Thus, the machinery of the step bunching in the case of SD bias for crystal growth (sublimation) process is well understood in the case when the adatom concentration $c_0$ is low (high). Then the question arises: why high density of adatoms for growth and high vacancy density for sublimation destroys the tendency for step bunching. The reasoning here is as follows: high (low) $c_0$ for growth (sublimation) means that steps move faster and this, combined with the low diffusion rate ($n_{DS}$ close to 1) causes that length of the terraces does not count and no step-step correlation is possible.

The mechanism of step bunching caused by SU bias is more complicated. It happens at high (low) adatom concentration for crystal growth (sublimation) and for high diffusion rate, $n_{DS} \gg 1$, that causes the steps to become transparent, because adatoms bypass them during long diffusional pathways. But even then the condition for large $n_{DS}$ is not enough to induce the bunching process but only surface roughening results. Our study shows that it is necessary to slow down the step motion by making it reversible - increasing the probability of sublimation during the growth process and the probability of growth during the sublimation process. Figure 1c and Figure 1d are obtained under such conditions. The scenario of the bunching instability in such a case can be understood as a communication between steps by the induced by step motion excess of adatom vacancies (adatoms) produced at one step, then transported towards the next one. When velocities of steps and adatoms are correlated speeding up of one steps slows down the next one and as a result step bunching occurs. The development of the instability in the SU case makes it more and more similar to the SD case at the same stage – larger terraces amplify the instability by the cumulation of the diffusional asymmetry over many jumps. In the same time, at both ends of an increasing terrace there are step formations with increasing collective impermeability – a step bunch of many permeable steps as a coarse grained object is itself increasingly impermeable proportionally to the number of steps in it. Thus the similarity between SD and SU cases increases making them indistinguishable in the intermediate stages of the instability and below we quantify this similarity by studying the time-scaling of the average number of steps and showing that it tends to the same, universal curve. The time-scaling of $N$ can be written in the general form as function of the rescaled time $T \equiv t/\tau$ as $N = aT^{\beta}$ and we find both the parameter combinations (characteristic time scale) $\tau$ and the time-scaling exponent $\beta$, as well as the numerical pre-factor (amplitude) $a$. The dependencies found for each of the four cases studied SD/Gr, SD/Sbl, SU/Gr, and SU/Sbl are shown in Figure 3. The curves were obtained after averaging over three runs for each set of parameters chosen. At first place, it is seen that in every case studied bunch size always achieves, soon or later, a power law dependence on the time with $\beta = 1/2$. This time-scaling exponent was obtained in many different studies, both theoretical/numerical and experimental, for the bunch size and for the terrace width, and may even seem trivial. We go beyond these by showing that the dependencies obtained in all four cases collapse onto a single master curve $N = 2\sqrt{T/3}$ and the parameter combinations $\tau$ used to rescale the time depend on the bias direction but do not depend on the vicinal process, growth or evaporation. The following two parameter combinations $\tau$ are used to re-scale the time:

$$\tau = \begin{cases} 4l_0/\delta n_{DS}, & SD\,bias \\ 2l_0/c_0(1-c_0)|\delta|n_{DS}, & SU\,bias \end{cases} \qquad (2)$$

Note that the time scale we use for step bunching in the SD biased case is slightly different than the one published in Ref.[56] (they both fail to collapse the data onto the master curve at $n_{DS}$>40 but still resulting in $\beta$=1/2), in the present study a numerical factor of 4 appears instead of $c_0$. For densities around 0.2, that were studied before, the difference is negligible. After studying a wider span of surface densities we have found that the proper choice of timescale for the SD case does not depend on the density. It can be also seen that both choices of timescales are symmetric for $c_0$ and (1-$c_0$) interchange what has its consequences in the reflexion symmetry of the diagrams present in Figure 1 and Figure 2.

We use equation (2) to estimate simulation time necessary to obtain credible stability diagrams – from one side, the evolution time is to be long enough to verify which set of parameters leads to emergence of SB and which does not, and, on the other hand, it should not be too long, thus permitting to perform quite more runs with different parameters. We assumed rescaled time $T_{gr}$ =500, thus $N \approx 25$, and calculated corresponding number of real (not rescaled) time steps $t_{gr}$ for each parameter combination on the grid.

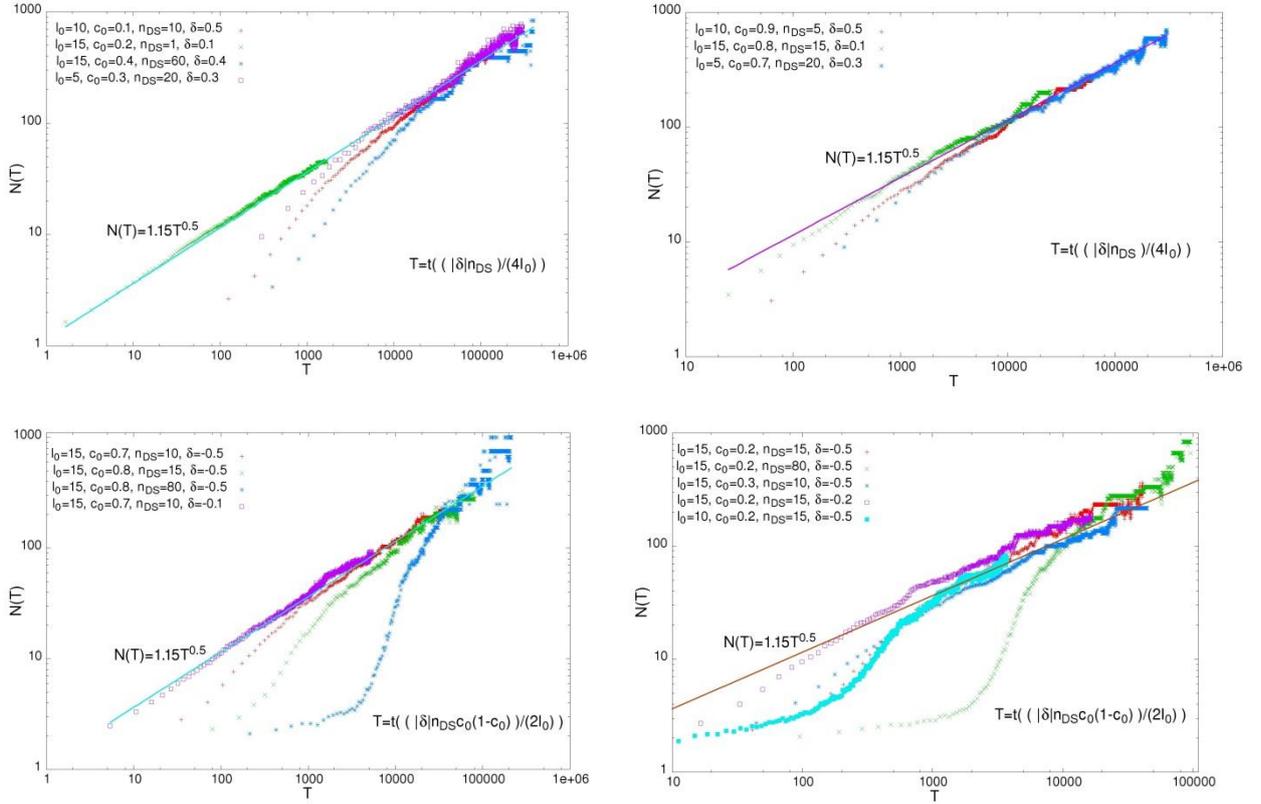

Figure 3. Time dependent bunch size for the four studied cases and different simulation parameters. Scaling parameters used (for growth and sublimation) differ for SD and SU cases, see equation (2) . Note that here we go well beyond the value of $T$=500 used to plot the stability diagrams.

In what follows we provide a perspective on the just obtained results for the time-scaling of the bunch size from vicCA in order to justify the numerical pre-factor. For this purpose, two models are studied based on systems of ordinary differential equations – one for the velocity of each step in the step train (ODE-system based models). The first one, that of Liu and Weeks (LW)[70] is aimed at the KL regime of the instability and is defined through its equation(s) for the step velocity $v_i$ (see also the Supplement) as:

$$v_i \equiv \frac{dx_i}{dt} = \Delta x_i + B \Delta x_{i+1} + R_i \qquad (3)$$

where the interplay of the first two terms containing neighboring terrace widths $\Delta x_i \equiv x_i - x_{i-1}$, destabilizes the surface when the parameter $B$<1, compare with eq.(1). The destabilizing impact of the surface kinetics is opposed by the omnipresent step-step interactions that tend to equilibrate the lengths of the two terraces adjacent to the step. In a non-dimensionalized version of LW, see the Supplement, the term that accounts realistically the role of the step-step repulsions is $R_i \equiv 2F_i^{-(n+1)} - F_{i+1}^{-(n+1)} - F_{i-1}^{-(n+1)}$, $F_i^{-(n+1)} \equiv \Delta x_i^{-(n+1)} - \Delta x_{i+1}^{-(n+1)}$ and the canonical value of $n$, the power of the step-step distance $r$ in the step-step repulsions law $U \sim 1/r^n$, is 2[71], stemming mainly from elastic, at low temperatures, or entropic, at high temperature, sources. We confront the results from LW with an *ad hoc*, minimal model (MM)[46] in which the first two terms (the destabilizing feedback) are the same as in LW, plus a negative feedback opposing the destabilization but a simpler than in LW one - $R_i = F_i^{-(n+1)}$. Here we show results only for the one-sided versions of the two models, i.e. $B=0$, then the destabilizing part simplifies to $\Delta x_i$ and no parameter remains to enter eventually the time-scaling. The time-scaling of the bunch size $N$ in LW and MM is independent also of the choice of $n$:

$$N = \frac{\sigma}{\sqrt{3}} T^{1/\sigma} \qquad (4)$$

where a further conjecture is made - $\sigma$ is the "slope exponent" – the order of the curve "surface slope vs. spatial coordinate", see the Supplement. In LW $\sigma=2$, while in MM it is 1. Note that MM provides also a counter-example to the eventual triviality of $\beta = 1/2$.

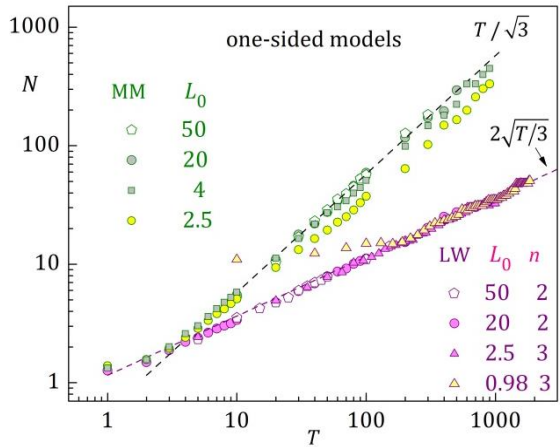
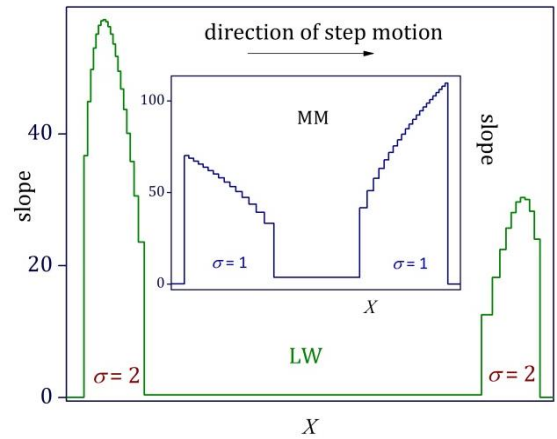

Figure 4. Time-scaling of the bunch size in the one-sided version of LW and MM models. Larger $L_0$ means here a larger destabilization. For MM $n = 2$, note also that when $L_0$ is small the data fail to collapse onto the universal curve, equation (4), but still preserve the exponent $\beta$.

Figure 5. Surface slope (~the derivative of the surface profile) as obtained from the two ODE-system based models – LW and MM. Two bunches are shown for each model before their merger. When not interacting, all bunches are oriented as the left one for each model. In LW the largest slope is in the middle of the bunches, in MM it is largest in the beginning of the bunches. In both models the slope increases with the bunch size $N$.

The time-scaling of $N$ is only one part of the whole scaling picture as predicted in [37] – in both LW/MM the SB phenomenon results in self-affine surfaces and two length-scales are necessary to describe them thoroughly[40]. Thus, while the bunch size $N$ corresponds to the vertical length-scale (height), an additional, horizontal length-scale enters the problem – the bunch width $W$ [37, 40]. In the case of our model vicCA, the lack of step-step repulsion to prevent the macrostep formation, we have shown[56] for the Gr+SD case that the role of a second lengthscale is played by the macrostep size $N_m$ and its time-scaling exponent $\beta_m$ distinguishes between the DL and KL regimes of the instability - $\beta_m \approx (3/4)\beta$ and $\beta_m = (3/5)\beta$, correspondingly (with $\beta=1/2$).

For LW, which corresponds to the assumption for KL regime of sublimation, we show in the Supplement that:

$$W \sim T^{1/z}; \quad 1/z = (n-1)/2(n+1) \tag{5}$$

with the important exception that for $n=2$ the exponent $1/z$ is $1/5$ instead of $1/6$ as follows from eq. (5). The exponent in eq. (5), together with the exponent of $N$, $\beta=1/2$, corresponds to the prediction of Pimpinelli-Tonchev-Videcoq-Vladimirova (PTVV)[37] for the $\rho=-1$ universality class but with a shift in the values of $n$. With such a shift the exponents in the time-scaing of the bunch width predicted by PTVV for values of $n''$ are obtained from the LW model with values of $n = n''+2$ (remember also the exception of the time-scaling exponent of the bunch width $1/5$ instead of $1/6$ as predicted formally by PTVV for $n'=0$). Thus the program started by PTVV [37] and modified by Krug et al.[64] by introducing a correction $k$ in order to distinguish between the DL and KL modes of the instability and transforming $n'$ into $n''=n'-k$, with values of $k$ being 0 (DL) and 1(KL), should be modified further – our study shows that for the LW-model (KL) $k=2$ in general (remember that $1/z=1/5$ when $n=2$!).

In conclusion, we have shown that vicCA model is able to reproduce the instability caused by the two opposite drift directions in the two fundamental situations of step motion – vicinal sublimation and vicinal growth. By performing a cross-validation between vicCA and LW model based on the time-scaling of $N$, the universality of intermediate stages of bunch dynamics is revealed. Our study opens up pathways for further studies in at least two general directions – further numerical study of the model of Sato and Uwaha [51] which permits by varying a single parameter, with a role similar to $n_{DS}$ as used here, to study both regimes of the instability, KL and DL, and, from the other side, an extension of vicCA in (2+1) D, thus studying and quantifying from a unified viewpoint the step bunching + step meandering instability. Then, more complex models could be revisited [47] that study the subtle interplay between the stabilizing and destabilizing factors.

The present research is supported by the National Science Centre (NCN) of Poland (Grant No. 2013/11/D/ST3/02700). VT, HP and AK acknowledge partial financial support from T02-8/121214 with the Bulgarian National Science Fund. Most of the calculations were done on HPC facility Nestum (BG161PO003-1.2.05).

**Supplement to:**

**Unstable dynamics of model vicinal surfaces:**

**Initial and intermediate stages**

F. Krzyżewski, M. Załuska-Kotur, A. Krasteva, H. Popova and V. Tonchev

I. Details on the atomistic scale model of unstable vicinal growth/sublimation (vicCA)

I.1 Parameter choice

In the case of SD bias applied along surface, bunching process happens at this part of the parameter space where naturally the rates of step motion are low. This is part where densities are small (high) in the case of growth (sublimation). We checked that for SD drift step bunching does not occur in the regions of high (small) in the case of growth (sublimation) even if we slow down the process by manipulating the sticking/detachment probabilities. The situation is opposite for the case of SU drift, where in order to get step bunching one needs to slow down the process of growth/sublimation. Thus both attachment and detachment of adatoms at the steps are realized at each simulation step. Moreover, it is very important that holes/adatoms produced at one step can reach the neighbouring step, otherwise no step bunching is observed. For given probabilities $p_G$ and $p_S$ the direction of step motion depends on the concentration of adatoms $c_0$. When it is large enough crystal grows because many adatoms attach to the steps and, on the other hand, detachment process is blocked by adatoms that occupy sites on top of the steps. When $c_0$ is small less adatoms are incorporated into the crystal, the sublimation becomes more frequent because sites on top of the steps are no longer blocked by the adatoms. Hence, in order to provide that crystal grows or sublimates for all density values we have to choose the adatom attachment and detachment probabilities properly. In order to study growing system we set attachment probability to be $p_G = 1-c_o+0.1$ and detachment probability equal to the density $p_S = c_0$. Single time unit of the simulation starts with attachment process, followed by $n_{DS}$ diffusional jumps. After that adatom detachment procedure was called and finally adatom concentration was updated in order to keep it constant and equal $c_0$. When the system is sublimating under SU bias we set step detachment probability as $p_S = c_o + 0.1$, attachment probability as $p_G = 1-c_o$, and we construct time unit by executing first detachment of adatoms, then $n_{DS}$ diffusional jumps, attachment at steps, and finally density compensation.

I.2 Growth rate

In Figure S1 we show how the growth rate $v_{gr}$ defined as $v_{gr} = n_{gr}/t_{gr}$ depends on the adatom concentration $c_0$. In this definition $n_{gr}$ is the number of mono-atomic layers grown for time $t_{gr}$. When $n_{DS}$ is low and when the adatom concentration reaches (from below) value of about 0.5, $v_{gr}$ becomes greater than 0.02 layers/unit time and the step bunching process is blocked (Figure 1a in the main text).

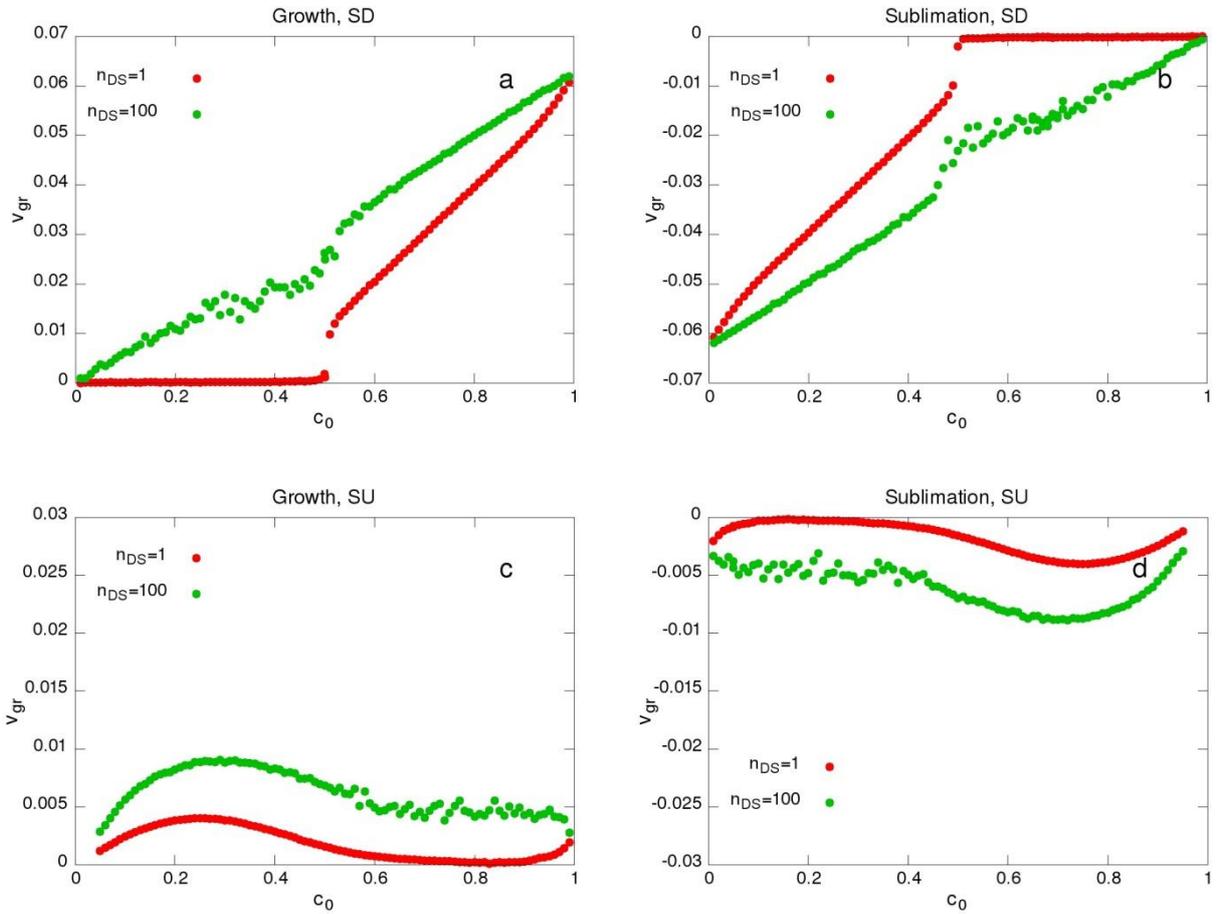

Figure S1 Concentration dependent growth rate in KL and DL regime at four investigated cases: growth at SD bias (a), sublimation at SD bias (b), growth at SU bias (c) and sublimation at SU bias (d).

Moreover, concentration dependent growth rate curves at DL and KL regime differ. When $n_{DS}$=100 (KL regime) $v_{gr}$-$c_0$ curve is almost linear whereas for low $n_{DS}$ values the character of analysed curves changes into non-linear one. Growth rate is low and almost constant for small densities of adsorbed adatoms and then increases rapidly for densities higher than 0.5. Above this critical value the number of sites occupied by adatoms is larger than number of empty sites. Diffusion of holes (i.e. empty lattice sites) rather than that of adatoms is realized effectively. At the same time more adatoms are consumed at steps and more new adatoms are added during the compensation stage. Positions of the adatoms become not so well defined. In the KL regime positions of adatoms, due to the fast diffusion, are "spread" also at low concentrations and growth rate increases as the probability of adatom encountered at steps increases. It can be seen however that in the region where step bunching occurs the rate of crystal growth is slightly reduced.

Similar behaviour with (1-$c_0$) playing role of $c_0$ is observed at the sublimated system with SD bias. Stability diagram for that case is presented in Fig. 1b of the main text. It is a mirror image (with respect to the y-axis) of the diagram for the growth+SD case, Fig 1a. The only kinetic process here is the detachment from the steps. Detachment probability is $p_S = 1$ and the adatoms do not attach back to the steps because $p_G = 0$. Thus, the negative values of the growth rate in Figure S1b mean sublimation. The system evolution is slower at high adatom concentration. It means slow detachment which is blocked by the presence of adatoms on top of the sites of the steps. The decrease of $c_0$ (increase of (1-$c_0$) ) leads to a faster sublimation because less adatoms are present at the surface and adatom detachment from steps happens more often. The departure from the DL regime towards a KL one changes the shape of the sublimation rate curve plotted as a function of concentration. In the case when $n_{DS} \gg 1$ the dependence is almost linear and a nonlinearity emerges when $n_{DS}$ decreases. As mentioned above, in the case of Figure S1b, system evolution is slower when concentration is higher. The stability diagram for this sublimation process shows that SB is strongest at low $n_{DS}$ values and concentrations greater than 0.4. It can be also summarized that low values of $n_{DS}$ (below 40) are the optimal regime for step bunching in both growth and sublimation processes destabilized by SD drift.

In Figure S1, c and d, one can see how growth and sublimation rate in the presence of SU drift behave as a function of the density. Again two different values of $n_{DS}$ are shown. Note that scale of the velocity in Figure S1, c and d, is two times smaller than in Figs. S1a and S1b., what means that systems grow and sublimate slowly. The rate of both processes is smaller in the bunching region and grows slightly when system evolves keeping the steps regularly distributed. Similarly to the case of SD drift the growth/sublimation rate increases when $n_{DS}$ is larger.

I.3 Bunch shape

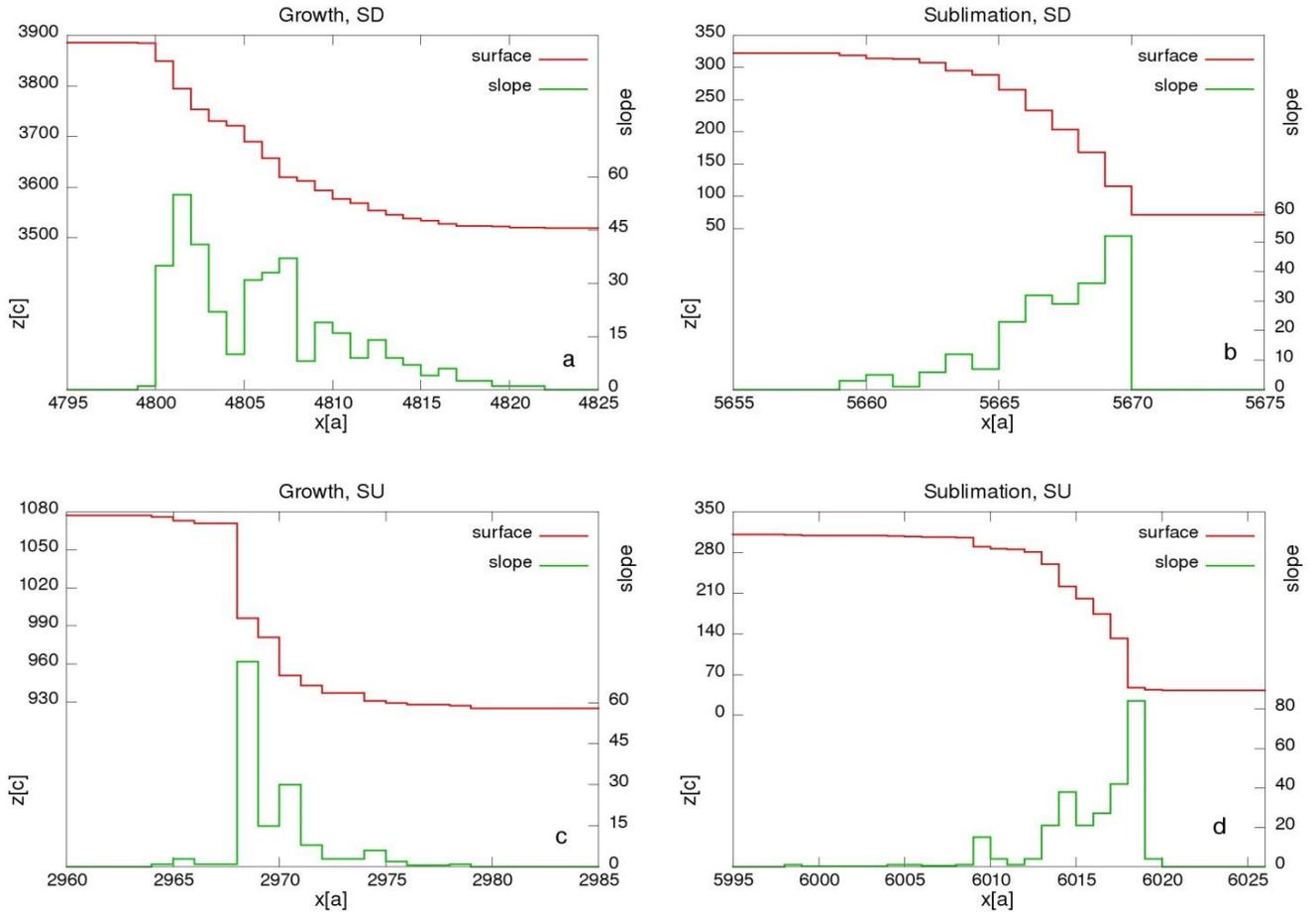

Figure S2. Bunches and their slopes observed at four investigated cases: growth at SD bias (a), sublimation at SD bias (b), growth at SU bias (c) and sublimation at SU bias (d).

Examples of bunches formed in the vicCA simulations are presented in Figure S2 where we show selected parts of surfaces and their slopes obtained during simulations. Specifically, the surface slope of the macrosteps is defined as $N_m/W_+$ where $W_+$ is the width of terrace below investigated macrostep of height $N_m$. Here one can see that bunch shapes differ. When the crystals are grown at SD bias or SU bias (Figure S2a and Figure S2c) bunches are steeper at higher (leftmost) parts. After sublimation with both directions of the bias (Figure S2b and Figure S2d) bunches are steeper at their lower (rightmost) areas. The reason for this is hidden in the direction of step motion. In the case of growth they move from the left side of the plots to the right. When system is sublimating the motion direction is opposite. Hence we can conclude that the front part (trailing edge) of each bunch is less steep than the back. The overall shapes of the bunches for SD and SU bias look similar, however there is a difference in the position of the steepest part of the bunch. For the SD biased systems it is always at the back edge of bunch, while for SU biased bunches it is placed inside, although close to the back edge. The difference in the shape is caused by the different process of formation of bunch structures.

I.4 Step trajectories

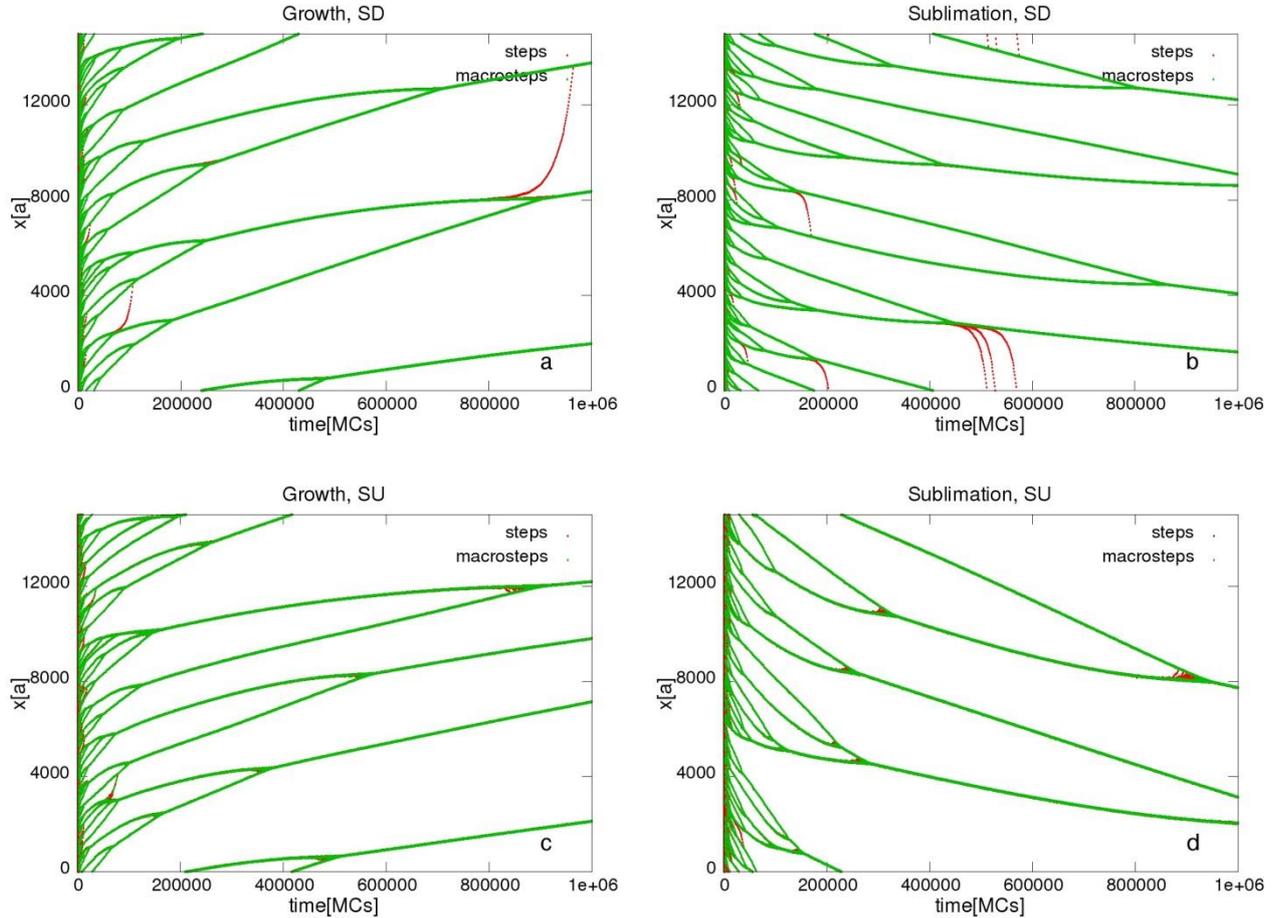

Figure S3 Step and macrostep trajectories observed at four investigated cases: growth at SD bias ($l_0$=10, $c_0$=0.1, $n_{DS}$=10, $\delta$=0.5) (a), sublimation at SD bias ($l_0$=10, $c_0$=0.9, $n_{DS}$=5, $\delta$=0.5) (b), growth at SU bias ($l_0$=15, $c_0$=0.7, $n_{DS}$=10, $\delta$=−0.5) (c) and sublimation at SU bias ($l_0$=15, $c_0$=0.2, $n_{DS}$=15, $\delta$=−0.5) (d).

In order to compare both scenarios of step bunching we can compare also the step trajectories. Figure S3 shows step (red lines) and macrostep (green lines) trajectories obtained from the four investigated cases of model evolution. In all plots macrosteps dominate the system's behavior. Single steps are very rare. They are faster than macrosteps and wander between them mediating a step exchange mechanism between neighbouring bunches. It is easy to see that the mechanism of bunching induced by SU bias (Figure S3, c and d) is rather weak and much less effective than this induced by SD bias (Figs S3a and S3b). In the Figure S3, a and b (SD bias), single steps detach from bunches at random moments of time and wander fast to the next ones, while in the system grown/sublimated with SU current (Figure S3, c and d) single steps are visible only at the moments preceding bunch connections. When surface is sublimated under SU bias single steps are very rare and almost absent. This observation can be interrelated to the fact that the bunching mechanism for SD bias is based on absorbing all adatoms or holes from the terrace, hence terraces are important, while for SU bias steps communicate between themselves. Hence in this last case bunches grow by connection of two neighbouring bunches together.

II. Liu and Weeks (LW) model

In this part of the supplement we extend our numerical study of the model of Liu and Weeks (LW) [70]. In their theoretical study of the sublimation Si(111) vicinal crystals controlled by the slow attachment/detachment rate of the adatoms to/from the steps they deduce a BCF-type equation (with non-transparent steps) for the velocity of a step in the step train:

$$\frac{dx_i}{dt} = \left(K_- \Delta x_i + K_+ \Delta x_{i+1}\right) + \frac{Kc_0 \Omega}{2kT} n g h^{n+1} \Omega \left[ 3\left(\frac{1}{\Delta x_i^{n+1}} - \frac{1}{\Delta x_{i+1}^{n+1}}\right) + \frac{1}{\Delta x_{i+2}^{n+1}} - \frac{1}{\Delta x_{i-1}^{n+1}} \right] \quad (6)$$

The first term is linear in the terrace widths $\Delta x_i \equiv x_i - x_{i-1}$ and when $K_- > K_+$ leads to destabilization of the initially uniform step train, $K_\pm \equiv \pm \frac{Kc_0 \Omega F}{2kT} + \frac{1}{2\tau_e}$, where $K$ is the attachment/detachment rate, $F$ is the electromigration force, only when negative it destabilizes a sublimating vicinal surface descending to the left, i.e step-down (SD), $c_0$ is the equilibrium concentration of the adatoms and $\tau_e$ is the average time the adatoms spend on the surface before evaporating in the ambience, $\Omega$ is the area occupied by an adatom. The second term comes from taking into account [72] the omnipresent step-step repulsions with magnitude of the repulsion energy $g$ [38] and $h$ is the height of a monoatomic step, the canonical value of the step-step repulsions exponent $n$ is 2. The coefficients $K_\pm$ are simplified further by introducing the dimensionless quantity $b \equiv -\frac{Kc_0 \Omega F \tau_e}{kT}$:

$$K_\pm \equiv \pm \frac{Kc_e \Omega F}{2kT} + \frac{1}{2\tau_e} = \frac{1}{\tau_e}\left(\frac{\pm Kc_e \Omega F/kT + 1}{2}\right) \equiv \frac{1}{\tau_e}\left(\frac{\mp b + 1}{2}\right) \quad (7)$$

to arrive at:

$$\frac{dx_i}{dt} = \frac{1+b}{2\tau_e}\left(B\Delta x_{i+1} + \Delta x_i\right) + u\left[3\left(\frac{1}{\Delta x_i^{n+1}} - \frac{1}{\Delta x_{i+1}^{n+1}}\right) + \frac{1}{\Delta x_{i+2}^{n+1}} - \frac{1}{\Delta x_{i-1}^{n+1}}\right] \quad (8)$$

where $u \equiv \dfrac{Kc_0\Omega}{2kT} ngh^{n+1}\Omega$. Now we non-dimensionalize (scale) eq. (8) by introducing the dimensionless coordinate and time $X \equiv \dfrac{x}{\xi}$ and $T \equiv \dfrac{t}{\tau}$:

$$\frac{dX_i}{dT} = \frac{\tau(1+b)}{2\tau_e}\left(\Delta X_i + B\Delta X_{i+1}\right) + \frac{\tau}{\xi^{n+2}}u\left[3\left(\frac{1}{\Delta X_i^{n+1}} - \frac{1}{\Delta X_{i+1}^{n+1}}\right) + \frac{1}{\Delta X_{i+2}^{n+1}} - \frac{1}{\Delta X_{i-1}^{n+1}}\right] \quad (9)$$

Since the scales for length and time $\xi$ and $\tau$ are arbitrary we can set the coefficients in front of the two terms on the rhs of eq. (9) equal to unity to obtain:

$$\frac{\tau(1+b)}{2\tau_e} = 1 \Rightarrow \tau \equiv \frac{2\tau_e}{(1+b)} \quad (10)$$

$$\frac{\tau}{\xi^{n+2}}u = 1 \Rightarrow \xi \equiv \left[\frac{2\tau_e}{(1+b)}u\right]^{1/(n+2)} \quad (11)$$

Thus the dimensionless version of the equation of LW model is obtained as:

$$\frac{dX_i}{dT} = \left(\Delta X_i + B\Delta X_{i+1}\right) + \left[3\left(\frac{1}{\Delta X_i^{n+1}} - \frac{1}{\Delta X_{i+1}^{n+1}}\right) + \frac{1}{\Delta X_{i+2}^{n+1}} - \frac{1}{\Delta X_{i-1}^{n+1}}\right] \quad (12)$$

Together with the system of equations (6) is defined the initial vicinal distance $\Delta x_i = l_0$, one for all values of $i$. It is non-dimensionalized to:

$$L_0 \equiv \left[\frac{2\tau_e}{(1+b)}u\right]^{-1/(n+2)} l_0 \quad (13)$$

More details, and especially how the procedure above can be used to find the scaling pre-factors, can be found in [73].

In its classical formulation, where $l_0$ and $\tau_e$ are used as scales for non-dimensionalization, the equaton of LW model is written as:

$$dx_i/dt = \Delta x_i(1+b)/2 + \Delta x_{i+1}(1-b)/2 + u'\left[3\left(\Delta x_i^{-(n+1)} - \Delta x_{i+1}^{-(n+1)}\right) + \Delta x_{i+2}^{-(n+1)} - \Delta x_{i-1}^{-(n+1)}\right] \quad (14)$$

where $u' = u\tau_e l_0^{n+2}$ and, hence, the connection between the two variants of non-dimensionalization is given by:

$$L_0 = \left[\frac{2u'}{(1+b)}\right]^{-1/(n+2)} \quad (15)$$

and

$$B = \frac{1-b}{1+b} \quad (16)$$

The LW model was studied further [41, 42, 64, 74] mainly with focus on the scaling relation between the minimal distance in the bunch $l_{min}$ (maximal slope) and the number of steps in the bunch, $l_{min} \sim N^{-\gamma}$ [75] (for a review see [76]). Unfortunately, the size scaling exponent $\gamma$ cannot distinguish between the DL and KL regime of the instability, as shown by Krug et al.[64]

Here we study in parallel the two length-scales, bunch size $N$ and bunch width $W$, necessary to describe thoroughly[37, 40] the self-affine patterns formed during the intermediate regime of the instability by using a unified monitoring protocol [39]. Thus we obtain the time-scaling exponents of the bunch size $N$ and of the bunch width $W$ by changing systematically the value of the step-step repulsions exponent $n$, see Figure S4 and Figure S5 below (the values of $L_0$ are shown on the left one only).

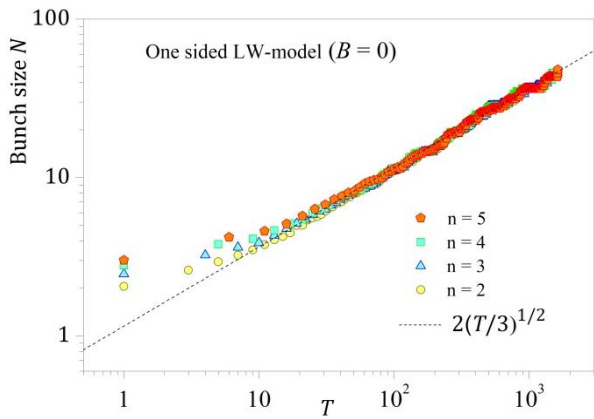 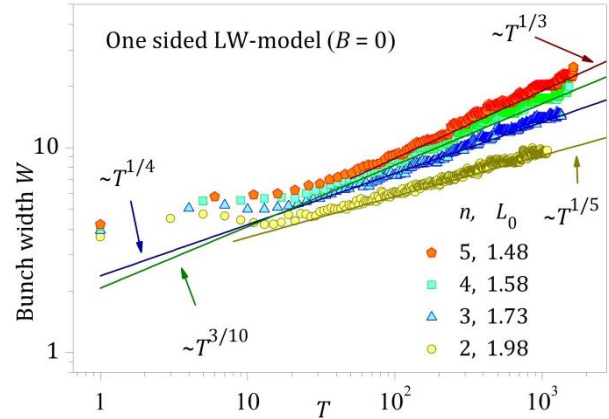

Figure S4 Average number of steps in the bunch $N$ as function of the rescaled time T from the one-sided case, $B=0$, of the LW model, the values of $L_0$ are given in the right panel. 3000 steps are used in this run.

Figure S5. Parallel results on the average bunch width as function of the rescaled time. It is the bunch width that distinguishes in between the values of the step-step repulsions exponent $n$.

III. Universality classes in step bunching (PTVV) [37, 64]

The hypothesis for the existence of universality classes in the SB phenomena was advanced by Pimpinelli et al. in [37](PTVV) based on a generalized version of a continuum equation for the height of the vicinal crystal surface constructed from two terms with opposite effects – destabilizing and stabilizing, they contain the generalizing exponents $\rho$ and $n$, respectively. Their equation is for the diffusion-limited (DL) regime of the phenomenon. Later, Krug et al., modified the equation in order to comprise also the KL regime by introducing a correction $k$ to the step-step repulsions exponent *n* -> *n-k* with values of *k* 0 or 1 for the DL/KL regime.

In order to help the reader in orienting through the "zoo" of exponents we provide below an extended table. The parallel study of the bunch size *N* and bunch width *W* in the one-sided version of the LW model as illustrated in Figure S4 and Figure S5 strongly supports the attribution of the LW model to the $\rho$=-1 universality class (although one could argue that also the universality class of the "C+ - C-" is possible based on the value of $\beta$) but with a the correction coefficient *k* as introduced by Krug et al. [64] should have the value of 2 for this, KL regime of the instability.

|  | C$^+$ - C$^-$ [39] (DL) | Universality Classes PTVV [37] | |
|---|---|---|---|
| **General *n*** | | | |
| $\alpha$ ($W \sim N^{1/\alpha}$) | (n+1)/n | (2+n-$\rho$)/(n-$\rho$) | |
| $\beta=\alpha/z$ ($N \sim t^\beta$) | 1/2 | (2+n-$\rho$)/[2(n+1-2$\rho$)] | |
| z ($W \sim t^{1/z}$) | 2(n+1)/n | 2(n+1-2$\rho$)/(n-$\rho$) | |
| $\delta$ ($l_b \sim t^\delta$) | 1/2(n+1) | 1/(n+1-2$\rho$) | |
| $\gamma$ ($l_b \sim N^{-\gamma}$) | 1/(n+1) | 2/(2+n-$\rho$) | |
| ***n*=1** | | $\rho$ = -2 | $\rho$ = -1 |
| $\alpha$ ($W \sim N^{1/\alpha}$) | 2 | 5/3 | 2 |
| $\beta=\alpha/z$ ($N \sim t^\beta$) | 1/2 | 5/12 | 1/2 |
| z ($W \sim t^{1/z}$) | 4 | 4 | 4 |
| $\delta$ ($l_b \sim t^\delta$) | 1/4 | 1/6 | 1/4 |
| $\gamma$ ($l_b \sim N^{-\gamma}$) | 1/2 | 2/5 | 1/2 |
| ***n*=2** | | $\rho$ = -2 | $\rho$ = -1 |
| $\alpha$ ($W \sim N^{1/\alpha}$) | 3/2 | 3/2 | 5/3 |
| $\beta=\alpha/z$ ($N \sim t^\beta$) | 1/2 | 3/7 | 1/2 |
| z ($W \sim t^{1/z}$) | 3 | 7/2 | 10/3 |
| $\delta$ ($l_b \sim t^\delta$) | 1/6 | 1/7 | 1/5 |
| $\gamma$ ($l_b \sim N^{-\gamma}$) | 1/3 | 1/3 | 2/5 |
| ***n*=3** | | $\rho$ = -2 | $\rho$ = -1 |
| $\alpha$ ($W \sim N^{1/\alpha}$) | 4/3 | 7/5 | 3/2 |
| $\beta=\alpha/z$ ($N \sim t^\beta$) | 1/2 | 7/16 | 1/2 |
| z ($W \sim t^{1/z}$) | 8/3 | 16/5 | 12/4 |
| $\delta$ ($l_b \sim t^\delta$) | 1/8 | 1/8 | 1/6 |
| $\gamma$ ($l_b \sim N^{-\gamma}$) | 1/4 | 2/7 | 1/3 |